\documentstyle[psfig]{mn}

\newif\ifAMStwofonts

\def\degr{\hbox{$^\circ$}}
\def\arcmin{\hbox{$^\prime$}}

\def\fs{\hbox{$.\!\!^{\rm s}$}}

\def\farcs{\hbox{$.\!\!^{\prime\prime}$}}

\newcommand{\lapp}{\mbox{\raisebox{-0.3em}{$\stackrel{\textstyle <}{\sim}$}}}
\newcommand{\gapp}{\mbox{\raisebox{-0.3em}{$\stackrel{\textstyle >}{\sim}$}}}

\title[Intermittent jet activity in 4C29.30?]
      {Intermittent jet activity in the radio galaxy 4C29.30?}
\author[M. Jamrozy et al.]
       {M. Jamrozy$^1$$\thanks{E-mail: jamrozy@oa.uj.edu.pl (MJ);
                   sckonar@ncra.tifr.res.in (CK);  
                   djs@ncra.tifr.res.in (DJS);
                   stawarz@oa.uj.edu.pl ({\L}S);
                   mack@ira.inaf.it (KHM); asiemiginowska@cfa.harvard.edu (AS)}$, 
        C. Konar$^2$, D.J. Saikia$^2$, {\L.} Stawarz$^{1,3,4}$, K.-H. Mack$^{5}$, A. Siemiginowska$^{6}$ \\
$^1$ Obserwatorium Astronomiczne, Uniwersytet Jagiello\'nski, ul. Or{\l}a 171, PL-30244 Krak\'ow, Poland \\
$^2$ National Centre for Radio Astrophysics, TIFR, Pune University Campus, Post Bag 3, Pune 411 007, India \\
$^3$ Kavli Institute for Particle Astrophysics and Cosmology, Stanford University, Stanford CA 94305, USA \\
$^4$ Stanford Linear Accelerator Center, Menlo Park CA 94025, USA \\
$^5$ Istituto Nazionale di Astrofisica, Istituto di Radioastronomia, Via P. Gobetti 101, I-40129 Bologna, Italy \\
$^6$ Harvard-Smithsonian Center for Astrophysics, 60 Garden Street, Cambridge, MA 02138, USA \\
}
\date{Accepted.                         Received }

\pagerange{\pageref{firstpage}--\pageref{lastpage}}
\pubyear{  }

\begin{document}

\maketitle

\label{firstpage}

\begin{abstract}
We present radio observations at frequencies ranging from 240 to 8460 MHz
of the radio galaxy 4C29.30 (J0840+2949) using the Giant Metrewave Radio Telescope 
(GMRT), the Very Large Array (VLA) and the Effelsberg telescope. We report the
existence of weak extended emission with an angular size of $\sim$520 arcsec (639 kpc)
within which a compact edge-brightened double-lobed source with a size of 
29 arcsec (36 kpc) is embedded. We determine the spectrum of the inner double
from 240 to 8460 MHz and show that it has a single power-law spectrum 
with a spectral index is $\sim$0.8. Its spectral age is estimated to be 
$\lapp$33 Myr. The extended diffuse emission has a steep spectrum with a spectral
index of $\sim$1.3 and a break frequency $\lapp$240 MHz. The spectral age is
$\gapp$200 Myr, suggesting that
the extended diffuse emission is due to an earlier cycle of activity. 
We reanalyse archival x-ray data from Chandra and 
suggest that the x-ray emission from the hotspots consists of a mixture of nonthermal 
and thermal components, the latter being possibly due to gas which is shock heated by the jets
from the host galaxy. 
\end{abstract}


\section{Introduction}
One of the interesting issues in our understanding of 
active galactic nuclei (AGN) is the duration of their active phase and
whether such activity is episodic. In radio galaxies and quasars the extended 
radio emission provides us with an opportunity to probe their history
via the structural and spectral information of the lobes.  A
striking example of episodic nuclear activity is when a new pair of radio lobes
is seen closer to the nucleus before the `old' and more distant radio
lobes have faded (e.g. Subrahmanyan, Saripalli \& Hunstead 1996; Lara et al. 1999).
Such sources have been christened as `double-double' radio galaxies
(DDRGs) by Schoenmakers et al. (2000). 
Approximately a dozen or so of such DDRGs are known in the literature (Saikia, 
Konar \& Kulkarni 2006, and references therein).

In addition, diffuse relic radio emission due to an earlier cycle of activity may 
also be visibile around radio sources which are not characterised by a `classical
double' structure with hotspots at the outer edges. 
The relic radio emission is expected to remain visible for $\sim$10$^8$ yr
or so (e.g. Owen, Eilek \& Kassim 2000; Kaiser \& Cotter 2002; Jamrozy et al. 2004),
and have a steep radio spectrum due to radiative losses.  Such radio
emission possibly due to an earlier cycle of activity has been suggested for a number of 
sources from both structural and spectral information. These include 3C338 and 3C388
(Burns, Schwendeman \& White 1983; Roettiger et al. 1994), 
Her A (Gizani \& Leahy 2003), 3C310 (van Breugel \& Fomalont 1984; 
Leahy et al. 1986), and Cen A (Burns, Feigelson \& Schreier 1983; Clarke, Burns \& Norman
1992; Junkes et al. 1993; Morganti et al. 1999).  
There have also been a few candidates amongst compact radio sources.  
For example, a lobe of emission seen on  one side of the nuclear region in the Giga-hertz 
Peaked Spectrum (GPS) source B0108+388 has been suggested to be a relic of a previous
cycle of jet activity (Baum et al. 1990), although the one-sidedness of the emission
is puzzling (cf. Stanghellini et al. 2005).  A search for small-scale halos, on scales 
larger than the known milliarcsec-scale structures of compact steep spectrum and GPS sources, 
using interplanetary scintillation observations with the Ooty Radio Telescope at 327 MHz 
led  to the identification of a few possible candidates (Jeyakumar et al. 2000). 
However, such features are not very common in either small or large radio sources. For example, 
a number of searches for such features 
have not yielded clear and striking examples of relic emission around bright radio sources 
(e.g. Reich et al. 1980; Stute, Reich \& Kalberla 1980; Perley, Fomalont \& Johnston 1982; 
Kronberg \& Reich 1983; van der Laan, Zieba \& Noordam 1984; Jones \& Preston 2001). 

\begin{figure*}
\hbox{
 \psfig{file=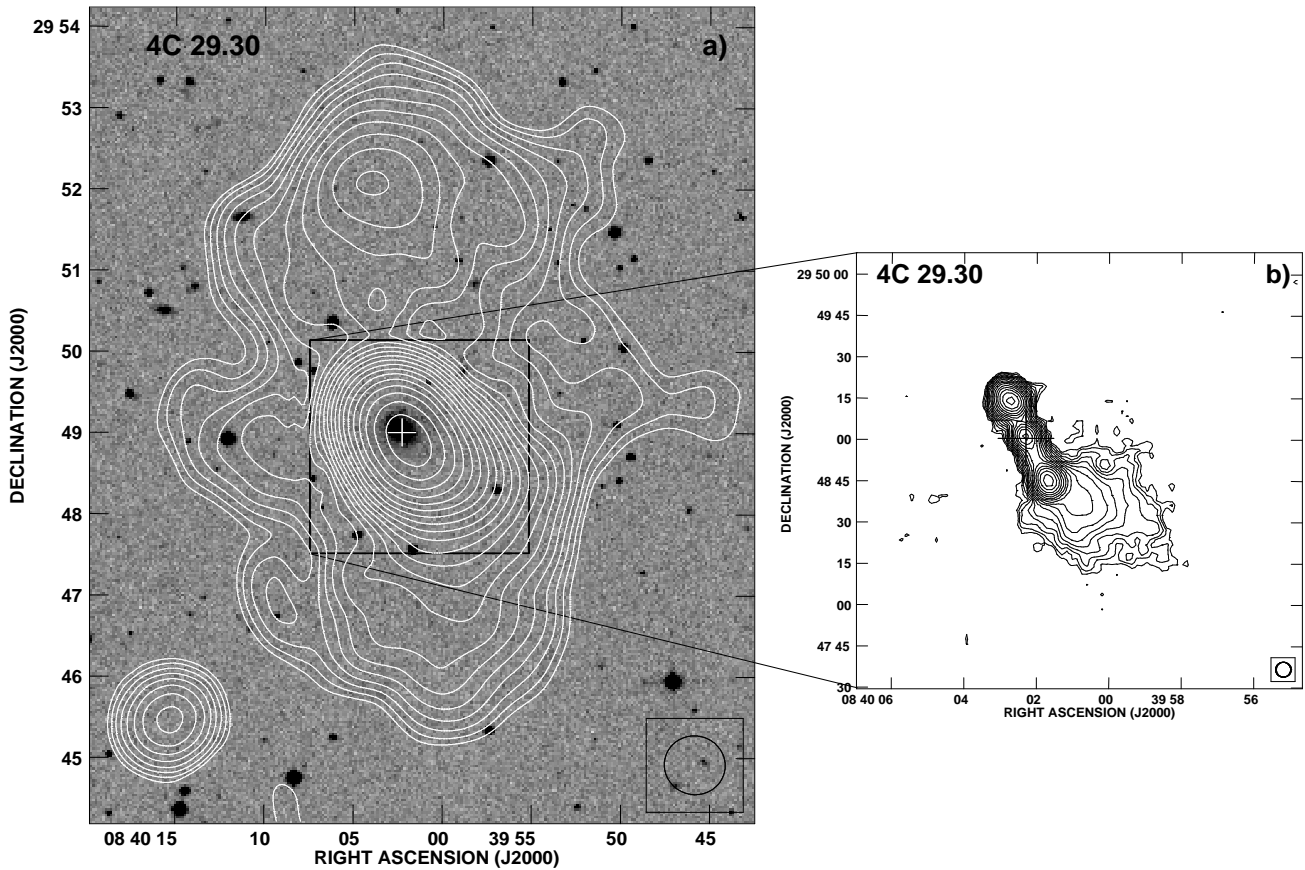,width=7.0in,angle=0}
}
\caption[]
{1400-MHz VLA images of 4C\,29.30. {\bf a)} D-array contour map of the entire source
overlayed on the optical field from the Digital Sky Survey (DSS). The contour levels
are spaced by factors of $\sqrt{2}$ and the first contour is 0.3~mJy~beam$^{-1}$.
{\bf b)} B-array contour map of the central part of the source from FIRST. The contour
levels are spaced by factors of $\sqrt{2}$, and the first contour is
0.45~mJy~beam$^{-1}$. The size of the beam is indicated by an ellipse in the
bottom right corner of each image. The cross marks the position of the radio core.}
\end{figure*}

In this paper we concentrate on the radio galaxy 4C29.30 (J0840+2949) which is associated with 
a bright (R $\rm\sim 15^{m}$) host elliptical galaxy (RA $\rm 08^{h}40^{m}02\fs370$,
DEC $+29\degr49\arcmin02\farcs60$; all positions being in J2000 co-ordinates throughout the paper) 
at a redshift of 0.06471$\pm$0.00013. The corresponding
luminosity distance is 287 Mpc and 1 arcsec corresponds to 1.228 kpc in a Universe with 
H$_\circ$=71 km s$^{-1}$ Mpc$^{-1}$, $\Omega_m$=0.27, $\Omega_\Lambda$=0.73 (Spergel et al. 2003). 
The published images show a double-lobed radio source which has two prominent
hotspots at the outer edges with an overall angular separation of 29 arcsec (36 kpc) 
and a prominent jet towards the south-west (e.g. van Breugel et al. 1986; Parma et al. 1986
and references therein). 
In addition there is a diffuse blob of emission towards the south-west (SW blob) with a size of
$\sim$40 arcsec (50 kpc) extending beyond the south-western hotspot. The radio luminosity 
of the inner double at 1400 MHz is 
5.5$\times$10$^{24}$ W Hz$^{-1}$,  which is significantly below the dividing line of the Fanaroff-Riley 
classes, while that of the entire source is 7.4$\times$10$^{24}$ W Hz$^{-1}$. It is interesting to 
note that in some of the DDRGs,
the luminosity of the inner double is in the FRI category although its structure resembles that of
FRII radio sources (cf. Saikia et al. 2006). 
A detailed radio and optical study of this galaxy (van Breugel et al. 1986)
shows optical line-emitting gas adjacent to the radio jet along a
position angle (PA) of $\sim$20$^\circ$ and evidence of the radio jets 
interacting with dense extranuclear gas. 

The host galaxy of 4C29.30 appears to have
merged with a gas-rich galaxy, shows presence of shells and dust
(Gonzalez-Serrano, Carballo \& Perez-Fournon 1993) and is associated with an IRAS source
F08369+2959 (Keel et al. 2005). At x-ray wavelengths Chandra detects  
emission from the hot spots in the southwestern radio lobe and also in the counterlobe 
(Gambill et al. 2003; Sambruna et al. 2004). Both hotspots have also been detected in
observations with the Hubble Space Telescope (Sambruna et al. 2004). 

In this paper we show the presence of diffuse extended emission on an angular
scale of $\sim$520 arcsec (639 kpc) which could be due to an earlier cycle of activity,
from observations with the Very Large Array (VLA), 
the Giant Metrewave Radio Telescope (GMRT) and the Effelsberg telescope.
This feature is also visible clearly in the Westerbork Northern Sky Survey (WENSS; 
Rengelink et al. 1997) at 325 MHz and the NRAO VLA Sky Survey (NVSS; Condon et al. 1998)
at 1400 MHz. We also present multifrequency observations of the inner double and the 
diffuse SW blob of emission with the GMRT and the VLA. This
should enable us to determine the spectrum more reliably and estimate the 
spectral ages and perhaps constrain the time scales of episodic activity in 
this source. For example, in the DDRG J1453+3308 we have been able to show using 
both GMRT and VLA observations that the spectrum of the outer lobes exhibits 
significant curvature while that of the inner lobes appears practically straight
(Konar et al. 2006). 

Our multifrequency observations and data reduction are
described in Section~2. The observational results, such as the radio maps showing the
source structure and spectra are presented in Section~3.  The results are presented in 
Section~4, while the concluding remarks are given in Section~5.

\section{Observations and data reduction}
The analysis presented in this paper is based on radio observations
made with the Effelsberg telescope, GMRT and VLA, as well as on VLA archival data.
The observing log for both the GMRT and VLA as well as Effelsberg observations is 
listed in Table~1 which is arranged as follows. Columns 1 and 2 show the name of the
telescope, and the array configuration for the VLA observations;
columns 3 and 4 show the frequency and bandwidth used in making the images;
column 5: the primary beamwidth in arcmin; column 6: dates of the observations.
The phase centre for all the observations was near the core of the radio
galaxy except for the VLA CnD array observations at 5 GHz where the antennas
were pointed about 2 arcmin north and south of the core in an effort to 
detect the diffuse extended emission.

\subsection{GMRT observations}
The observations were made in the standard manner, with  each observation
of the target-source interspersed with observations of 3C286 which was
used as a phase calibrator as well as flux density and bandpass calibrator.
At each frequency the source was observed in a
full-synthesis run of approximately 9 hours including calibration overheads.
The rms noise in the resulting full-resolution images ranges from about 
1 mJy beam$^{-1}$ at 240 MHz to about 0.06 mJy beam$^{-1}$ at 1287 MHz.
Details  about the array can be found at the GMRT website at
{\tt http://www.gmrt.ncra.tifr.res.in}. The data collected were calibrated 
and reduced in the standard way using the NRAO {\tt AIPS} software package. 
Several rounds of self calibration were done to improve the quality of the images.
The absolute position uncertainty could be up to several arcsec in the low-frequency
images due to phase errors introduced by the ionosphere (cf. Rengelink et al. 
1997). The flux densities at the different frequencies are consistent with the scale of
Baars et al. (1977). 

\begin{figure}
\begin{center}
\hspace{0.5cm}
 \psfig{file=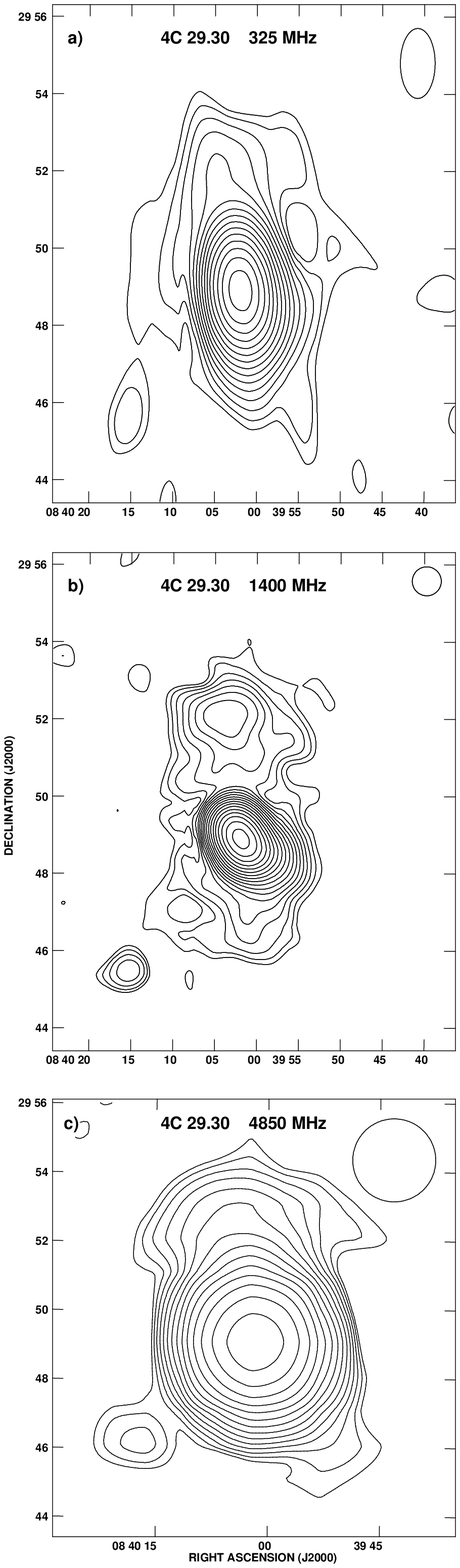,width=2.5in,angle=0}
\caption[]{Atlas of the radio source 4C\,29.30 at three different radio frequencies.
{\bf a)} 325-MHz map from the WENSS, {\bf b)} 1400-MHz map from the
NVSS, {\bf c)} 4850-MHz map from the  Effelsberg observations.
The contours spaced by factors of $\sqrt{2}$ in brightness are
plotted starting with 9, 0.9 and 1.8 mJy~beam$^{-1}$ respectively. 
The size of the beam is indicated by an ellipse in the top
right corner of each image.}
\end{center}
\end{figure}

\begin{table}
\caption{Observing log. }
\begin{tabular}{l c r c c c }
\hline
Teles-    & Array  & Obs.   & Band- & Primary &  Obs. Date  \\
cope      & Conf.  & Freq.  & width &  beam   &             \\
          &        &(MHz)   &(MHz)  &(arcmin) &             \\
 (1)      &  (2)   & (3)    & (4)   &  (5)    &    (6)      \\
\hline
GMRT      &        & 240    & 4.5   &  114    & 2005 Dec 28   \\
GMRT      &        & 333    & 12.5  &   81    & 2005 Dec 24   \\
GMRT      &        & 605    & 12.5  &   43    & 2005 Dec 28   \\
GMRT      &        & 1287   & 12.5  &   26    & 2005 Dec 22   \\
VLA$^a$   &    D   & 1400   &  25   &   30    & 2001 Nov 30   \\
VLA$^a$   &    D   & 1400   &  25   &   30    & 2001 Dec 24   \\
VLA$^a$   &    C   & 4860   &  50   &    9    & 1988 Apr 14    \\
VLA       &   CnD  & 4860   &  50   &    9    & 2005 Oct 9, 10 \\
VLA       &   CnD  & 8460   &  50   &   5.4   & 2005 Oct 12    \\
Effelsberg&        & 4850   & 500   &   2.1   & 2004 Jun 25    \\
\hline
\end{tabular}

$^a$: VLA archival data.
\end{table}

\subsection{New and archival VLA observations}
The source was observed with the CnD array at a frequency of 4860 
and 8460 MHz to determine the spectra over a large frequency range by 
combining these results with those from the 
low-frequency GMRT images.  The source was observed in the
snapshot mode, the integration time for each scan being $\sim$20 min.
At 4860 MHz the source was also observed with the phase centres 
shifted by $\sim$2 arcmin towards the north and south of the core to 
image the diffuse extended emission. The interferometric phases were 
calibrated with the phase calibrator J0830+241. The source 3C286 was 
used as the primary flux density calibrator. For the images produced from the data
sets where the phase centres have been shifted by $\sim$2 arcmin, correction 
for the primary beam pattern has been done. We also supplemented our observations
with VLA archival data to examine evidence of variability of the core and
determine the spectrum of the inner double-lobed source. These observations
were made with the C-array at 4860 MHz. 

We retrieved the L-band data of 4C29.30 from the VLA archive. The unpublished
observations (proposal number AL515) were carried out in the
D-array configuration using the `Correlator Mode 4' with a bandwidth of 25 MHz
and central frequencies of 1365 and 1435 MHz.  The observations were performed
in two runs on 2001 November 30 and December 24 with the pointing centre at
RA $\rm 08^{h}40^{m}02\fs3$, DEC $+29\degr49\arcmin03\farcs0$
and a total integration time of $2\times$50~min. 
The telescope gains were calibrated using the calibration source 3C147. 
The source J0741+312 was used as a
phase calibrator. After the initial data reduction the two data sets were merged
using the task DBCON. After preliminary CLEANing of the map with the
routine IMAGR, several self-calibrations were performed to improve its quality.
Finally, the map was corrected for primary beam attenuation. 
As in the case of the GMRT data, all the VLA data were edited and reduced using the 
{\tt AIPS} package.  All flux densities are on the Baars et al. (1977) scale.  

\subsection{Effelsberg observations}
The 4850-MHz observations of 4C\,29.30 were carried out on 2004 June 25 with the
2-horn receiver system (Thierbach, Klein \& Wielebinski 2003) at the secondary focus of the Effelsberg
100$-$m telescope. A total of 6 coverages, with a size of $20\times20$ arcmin$^2$ in azimuth
and elevation, were obtained. The scanning speed was 40 arcmin/min and the
scan separation 1 arcmin. Calibration and pointing performance of the system
were checked by mapping and cross-scanning the point-like sources, 3C286 and
B0851+20. We adopt the flux density scale of Baars et
al. (1977). The standard MPIfR-NOD2 (Haslam 1974) software package was used for the
data reduction. Firstly, all single-pixel noise spikes were removed
by hand from each of the scans. Then the individual maps were combined using
the PLAIT algorithm by Emerson \& Gr\"{a}ve (1988). The rms noise
of 0.9 mJy beam$^{-1}$ in total power was estimated from off-source regions of the image well
away from its edges and the half-power beam width was determined to be
144.5 arcsec. Finally, the NRAO {\tt AIPS} software package was used for 
imaging the source. 

\section{Observational results}
\subsection{Overall structure}
Our VLA image at 1400 MHz with an angular resolution of $\sim$45 arcsec (Fig. 1) shows the 
entire structure of 4C29.30 which has a largest angular extent of 520 arcsec (639 kpc), 
in addition to the inner structure with an angular size of 29 arcsec (36 kpc). The rms
noise in the image is 0.13 mJy beam$^{-1}$ while the total flux density is 756 mJy. There is
a peak of emission with a peak brightness of 10 mJy beam$^{-1}$ located at 
RA $\rm 08^{h}40^{m}04\fs03$, DEC $+29\degr52\arcmin04\farcs4$ in the northern region of
the extended emission which is brighter than the south which has a peak brightness
of $\sim$3.5 mJy beam$^{-1}$ at RA $\rm 08^{h}39^{m}59\fs29$, DEC $+29\degr46\arcmin32\farcs3$.
We have reproduced in Fig. 1 the FIRST (Faint Images of the Radio Sky at Twenty-cm,
Becker, White \& Helfand 1995) 
image of the source with an angular resolution of 5.4 arcsec which shows the inner 
double-lobed structure and the SW blob of emission. 
Due to lack of short baselines this VLA B-array survey is insensitive to the 
extended structures. 

\begin{figure*}
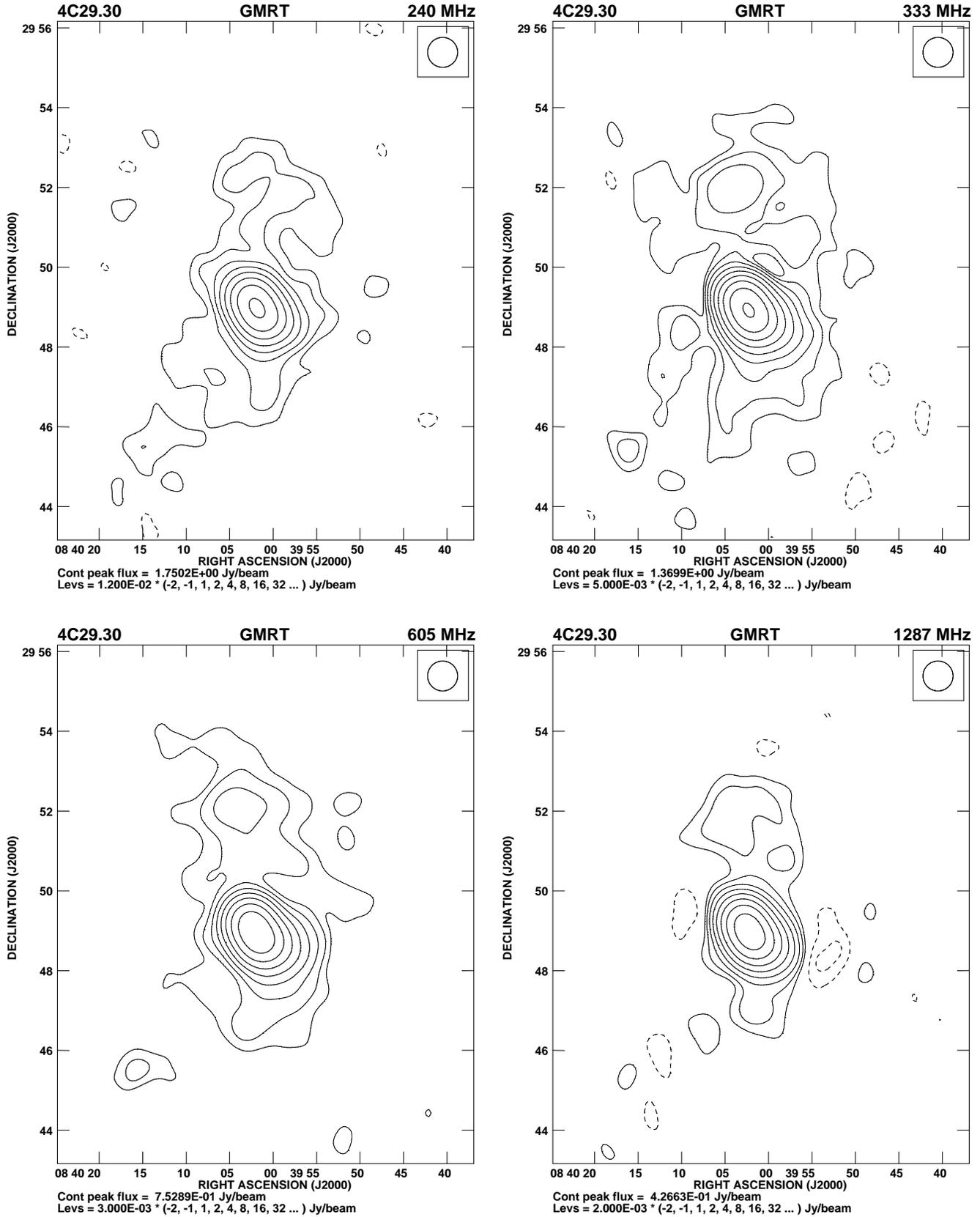

\vbox{
\hbox{
 \psfig{file=4C29.30_T_45arcsec.ps,width=3.5in,angle=0}
 \psfig{file=4C29.30_P_45arcsec.ps,width=3.5in,angle=0}
     }
\hbox{
 \psfig{file=4C29.30_G_45arcsec.ps,width=3.5in,angle=0}
 \psfig{file=4C29.30_L_45arcsec.ps,width=3.5in,angle=0}
     }
     }
\caption[]{The GMRT images of 4C29.30 at 240 MHz (upper left), 
333 MHz (upper right), 605 MHz (lower left) and 1287 MHz (lower
right). All these images have been made with an angular resolution
of 45 arcsec which is shown as a circle in the top right-hand
corner.}
\end{figure*}

\begin{figure*}
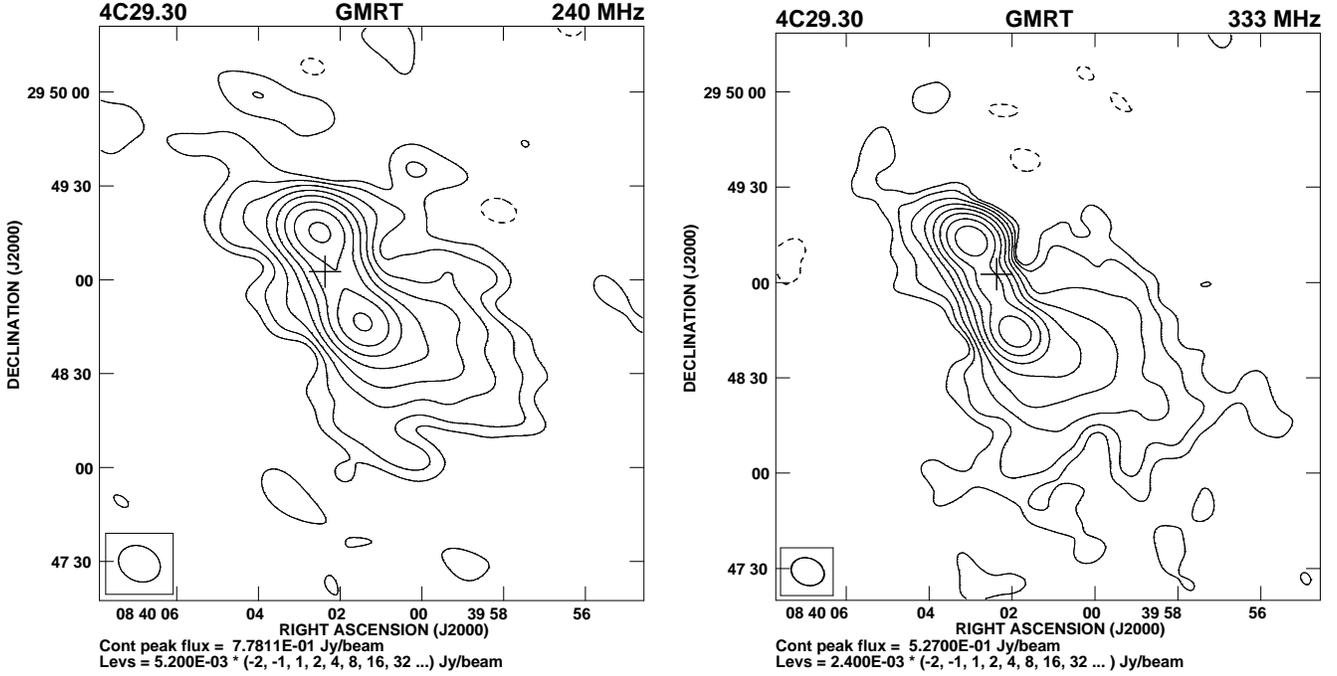

\hbox{
 \psfig{file=GM_4C29.30_T.ps,width=3.5in,angle=0}
 \psfig{file=GM_4C29.30_P.ps,width=3.5in,angle=0}
}
\caption[]{The GMRT images of 4C29.30. Left-hand panel: image at 240 MHz with 
an angular resolution of 13.8$\times$11.1 arcsec$^2$ along PA 64$^\circ$. 
Right-hand panel: the 333-MHz image with 
an angular resolution of 10.6$\times$8.5 arcsec$^2$ along PA 68$^\circ$. The
resolution is shown as an ellipse in the bottom left-hand corner while the 
$+$ sign marks the position of the optical galaxy.} 
\end{figure*}

\begin{figure*}
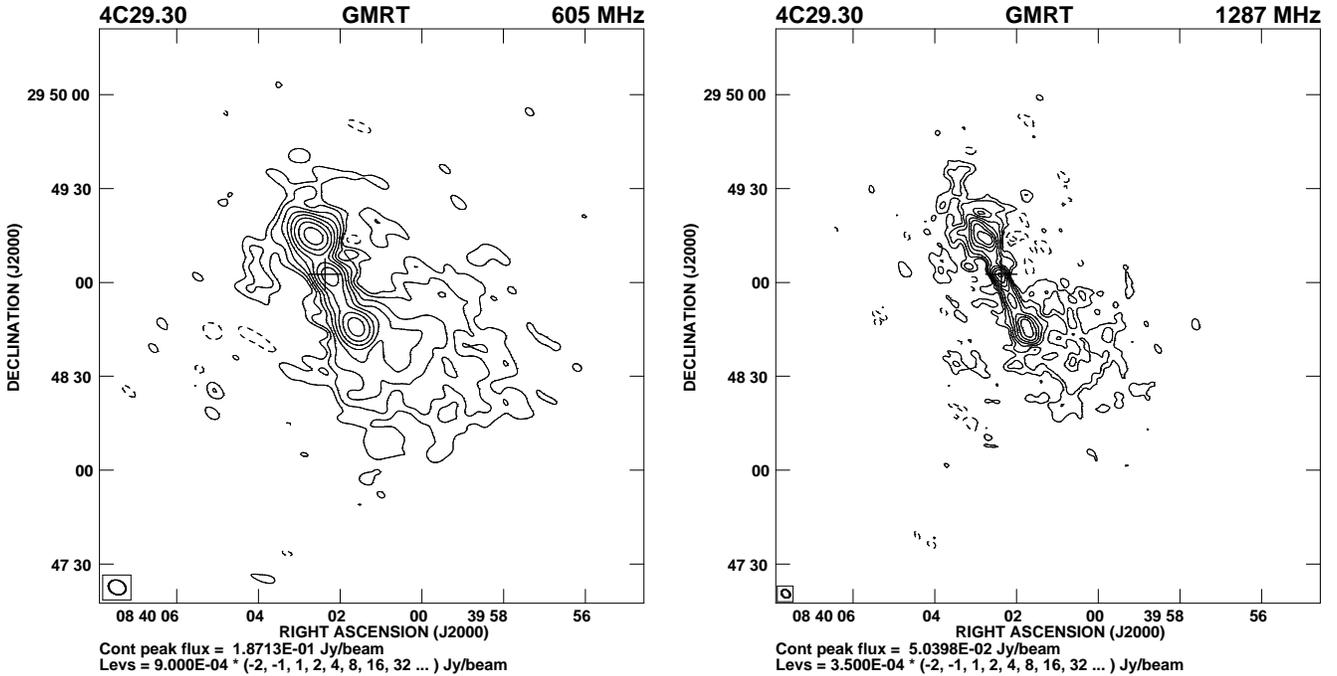

\hbox{
 \psfig{file=GM_4C29.30_G.ps,width=3.5in,angle=0}
 \psfig{file=GM_4C29.30_L.ps,width=3.5in,angle=0}
}
\caption[]{The GMRT images of 4C29.30. 
Left-hand panel: image at 605 MHz with 
an angular resolution of 5.66$\times$4.50 arcsec$^2$ along PA 66$^\circ$. 
Right-hand panel: image at 1287 MHz with an angular resolution of 
2.97$\times$2.30 arcsec$^2$ along PA 46$^\circ$. The
resolution is shown as an ellipse in the bottom left-hand corner while the 
$+$ sign marks the position of the optical galaxy.} 
\end{figure*}

\begin{figure*}
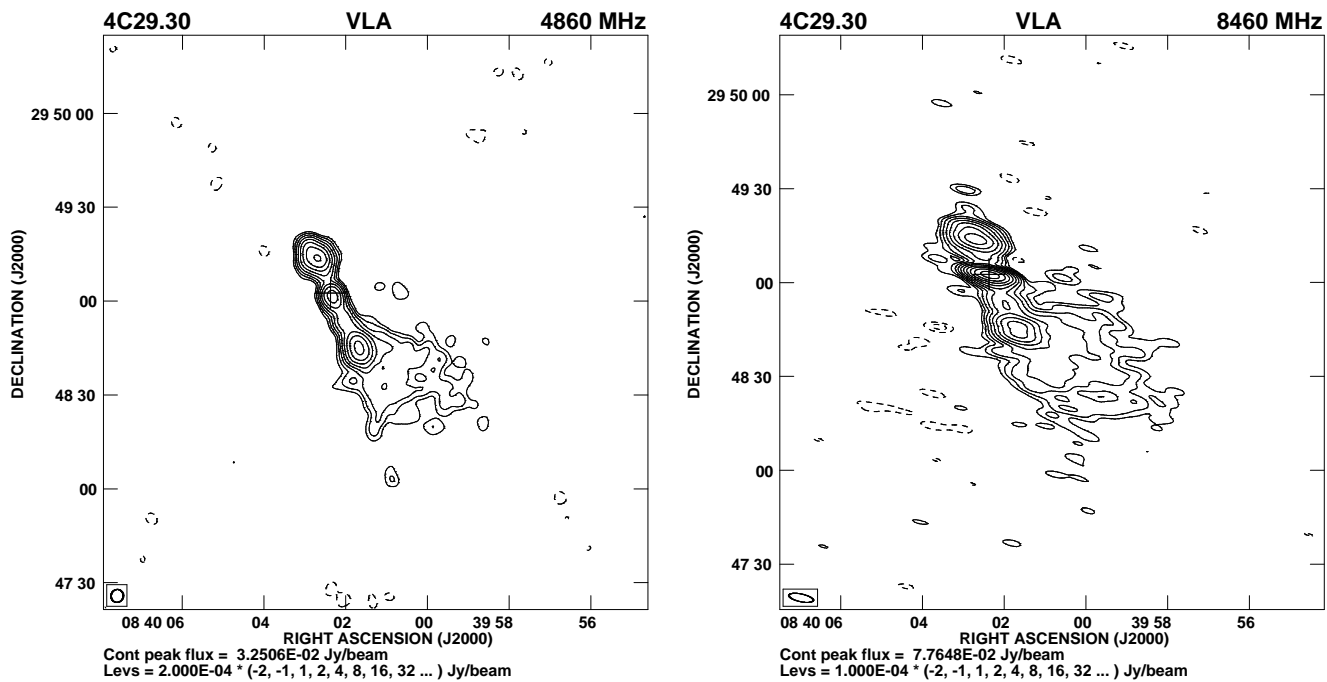

\hbox{
 \psfig{file=VLA.archiv_4C29.30_C.ps,width=3.5in,angle=0}
 \psfig{file=VLA.MJ_4C29.30_X.ps,width=3.5in,angle=0}
}
\caption[]{Left-hand panel: the VLA image of 4C29.30 at
4860 MHz with an angular resolution of
4.13$\times$3.90 arcsec$^2$ along PA 162$^\circ$. 
Right-hand panel: the VLA image at 
8460 MHz with an angular resolution of
7.90$\times$2.46 arcsec$^2$ along PA 78$^\circ$. 
The resolution is shown as an ellipse in the bottom left-hand corner while the
$+$ sign marks the position of the optical galaxy.}
\end{figure*}

The diffuse extended structure can also be seen in our Effelsberg image at 4850 MHz
as well as the WENSS image at 325 MHz and the NVSS image at 1400 MHz (Fig. 2). 
The 325-MHz  WENSS  map of the total intensity emission of
4C29.30 with the angular resolution of about $54\times109$ arcsec$^2$ (Fig. 2a)
has an rms noise of 3.9 mJy beam$^{-1}$ while the total flux density of the entire
source is 2170 mJy. The total intensity 1400-MHz VLA D-array NVSS map of the source with
an angular resolution of $45\times45$ arcsec$^2$ and an rms noise of 0.45 mJy beam$^{-1}$
(Fig. 2b) has a total flux density of 731 mJy, which is only marginally smaller than
the estimate of 756 mJy from our image (Fig. 1). 

Although only the inner double and the SW blob of emission are visible in the 
full-resolution GMRT images as discussed below, tapered images with an angular resolution 
of 45 arcsec were made from the
GMRT data at 240, 333, 605 and 1287 MHz to detect the diffuse extended
emission. These images are presented in Fig. 3 and have an rms noise 
of 4.6, 1.6, 0.9 and 0.5 mJy beam$^{-1}$ respectively,
while the total flux densities are 3541, 2647, 1465 and 727 mJy respectively. 
A comparison of the GMRT low-resolution images with our VLA D-array image (Fig. 1)
shows that although the northern peak and some diffuse emission have been detected,
weaker diffuse emission have been missed in these images. While at the lower
frequencies this is due to the higher rms noise values, the flux density at 1287 MHz
could also be affected by the lack of short spacings.

\subsection{The inner structure}
The inner structure of 4C29.30 has been observed several times at different
frequencies, viz. at $\sim$1400 MHz (Fanti et al. 1977, van Breugel et al. 1986,
Parma et al. 1986, Capetti et al. 1993),  $\sim$5000 MHz (Fanti et al. 1977,
van Breugel et al. 1986, Giovannini et al. 1988), and $\sim$15000 MHz (van Breugel
et al. 1986). The maps show two arrow-shaped lobes. The more compact, northern
lobe is located close to the radio core while the southern one, which is connected
with the nucleus by a distinct radio jet, is more extended. 
The inner double$-$lobed structure is oriented along a PA of $\sim$26$^\circ$
which is misaligned from the outer extended emission by  $\sim$18$^\circ$. The outer
emission is oriented close to that of the optical filament towards the south noted
by van Breugel et al. (1986). 

The GMRT low-frequency images at 240, 333, 605 and 1287 MHz 
which show the inner double and the SW blob of emission are shown in Figs. 4 and 5, 
while the VLA images at 4860 and 8460 MHz are presented in Fig. 6.
The flux densities estimated from these images are listed in Tables 2 and 3.
The observational parameters and some of the observed
properties are presented in Table 2, where we list the values estimated from the full-resolution
images as well as those estimated from the images made by tapering and weighting the data
to match the resolution of the GMRT image at 240 MHz. Table 2 is arranged as follows.
Column 1: frequency  of observations in MHz, with the letter G or V representing
either GMRT or VLA observations; columns 2$-$4: the major and minor axes of the restoring
beam in arcsec and its PA in degrees; column 5: the rms noise in units of 
mJy beam$^{-1}$; column 6: the peak flux density in the image in units of mJy beam$^{-1}$; 
column 7: the total flux density of only the inner double in units of mJy;
column 8: the total flux density of the inner double along with its
small extension to the south-west in units of mJy. 
The flux densities have been estimated by specifying a polygon around the source. 
The error in the flux density is approximately 15 per cent at 240 MHz and 7 per cent 
at the higher frequencies. 

\begin{table}
\caption{The observational parameters and flux densities of the inner structure.}
\begin{tabular}{l rrr c r r c}
\hline
Freq.        & \multicolumn{3}{c}{Beam size}   & rms & S$_p$  & S$_t^{id}$  &  S$_t^{id+SW}$    \\
(MHz)        & ($^{\prime\prime}$) & ($^{\prime\prime}$) & ($^\circ$) &    (mJy   & (mJy   &  (mJy)  & (mJy)   \\
             &         &       &       & /b)      & /b)   &       &          \\ 
(1)          & (2)     & (3)   & (4)   & (5)      & (6)   & (7)   & (8)      \\ 
\hline
  G240       & 13.8    & 11.1  &  64   &   1.58   &  778  & 2149  &   2623     \\

  G333       & 10.6    &  8.5  &  68   &   1.37   &  527  & 1680  &   2109     \\
             & 13.8    & 11.1  &  64   &   0.90   &  589  & 1805  &   2141     \\

  G605       & 5.66    & 4.50  &  66   &   0.37   &  187  &  891  &   1127     \\
             & 13.8    & 11.1  &  64   &   0.24   &  310  & 1046  &   1196     \\

  G1287      & 2.97    & 2.30  &  46   &   0.11   &  50   &  390  &    490     \\
             & 13.8    & 11.1  &  64   &   0.42   &  169  &  601  &    670     \\

  V4860      & 4.13    & 3.90  & 162   &   0.05   &  33   &  172  &    208     \\
             & 13.8    & 11.1  &  64   &   0.13   &  70   &  208  &    238     \\

  V4860      & 16.3    & 4.51  &  65   &   0.10   &  67   &  242  &    282     \\

  V4860      & 14.3    & 4.37  &  72   &   0.22   &  67   &  242  &    291     \\

  V4860      & 14.0    & 4.24  &  71   &   0.09   &  65   &  242  &    290     \\
  
  V8460      & 7.90    & 2.46  &  78   &   0.03   &  78   &  188  &    217     \\
             & 13.8    & 11.1  &  64   &   0.06   &  90   &  199  &    227     \\

\hline
\end{tabular}
\end{table}

\begin{table}
\caption{Flux densities of the radio core.}
\begin{tabular}{c l c r l c}
\hline
Telescope    &  Date of obs. & Freq. & Resn.      & Flux   & Refs.  \\
             &               &       &            & density$^a$&           \\
             &               & (MHz) & ($^{\prime\prime}$)  &        &          \\
  (1)        &    (2)        & (3)   &  (4)       & (5)    &  (6)   \\
\hline
 GMRT        & 2005 Dec 22   & 1286  & $\sim$2.6  &  50$^p$  (62$^p$) &  1      \\
 VLA-B       & 1993 Apr 02   & 1400  &       5.4  &  46$^p$  (46$^p$) &  2      \\
 VLA-A       & 1982 Feb      & 1452  &       1.3  &  51               &  3      \\
 VLA-B       & 1982 Aug, Sep & 1465  & $\sim$4.0  &  40               &  4      \\
 VLA-A       & 1982 Feb      & 4873  &       0.3  &  10.4             &  3      \\
 VLA-A       & 1985 Jan      & 4860  & $\sim$0.4  &   8.2             &  5      \\
 VLA-C       & 1988 Apr 14   & 4860  & $\sim$4.0  &  19$^p$ (21$^p$)  &  1      \\
 VLA-CnD     & 2005 Oct 9, 10& 4860  & $\sim$7.5  &  63$^p$ (66$^p$)  &  1      \\
 VLA-CnD     & 2005 Oct 12   & 8460  & $\sim$4.5  &  78$^p$ (81$^p$)  &  1      \\
\hline
\end{tabular}
\begin{flushleft}
$^a$: The core flux densities from our data and FIRST
are peak values in units of mJy beam$^{-1}$ estimated from two-dimensional Gaussian fits
to minimise contamination from extended emission. These have been marked with the 
superscript $^p$. The remaining values are from the literature and are in units of mJy.
The values within brackets are from images made with an angular resolution of 5.4
arcsec approximately along the jet axis so that the contamination from jet emission
is similar. \\
References. 1: Present paper. 
2: FIRST image; 
3: van Breugel et al. (1986). The flux density of C=C1+C2 
in their nomenclature
which would not be resolved by a $\sim$5 arcsec beam
is 16.5 mJy at 4873 MHz and 51 mJy at 1452 MHz, which is listed
in the Table here. 
4: Parma et al. (1986). 5: Giovannini et al. 1988. The flux
density of c+A+B in their nomenclature which would be 
unresolved by a $\sim$5 arcsec beam is 16 mJy.
\end{flushleft}
\end{table}

\begin{figure}
\vbox{
    \psfig{file=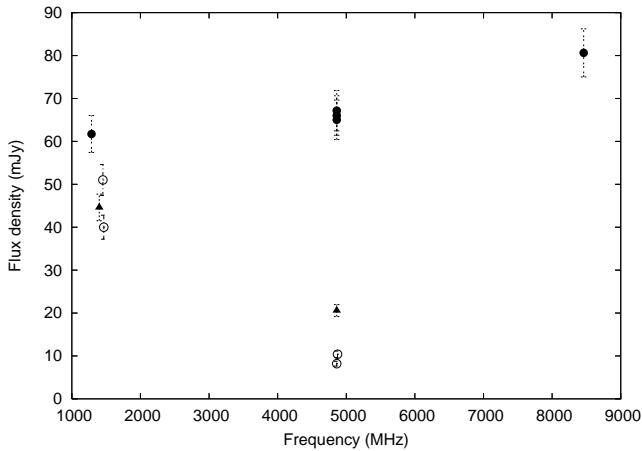,width=3.4in,angle=-90}
}
\caption[]{Spectrum of the core of 4C29.30. The measurements in
2005 with an angular resolution of 5.4 arcsec along the jet axis
are shown by filled circles, while earlier observations from 
FIRST and our analysis of archival VLA data with a
similar resolution are shown as filled triangles. The values from
the literature listed in Table 3 are shown as open circles.}
\end{figure}

\subsection{The radio core}
Our higher-resolution images (Figs. 5 and 6) clearly show the radio core in 
addition to the lobes of radio emission.
The J2000.0 position of the radio core estimated from our high-resolution image at
8460 MHz is RA $\rm 08^{h}40^{m}02\fs35$, DEC $+29\degr49\arcmin02\farcs5$,
which is consistent with the position of the optical galaxy 
(RA $\rm 08^{h}40^{m}02\fs370$, DEC $+29\degr49\arcmin02\farcs60$) from NED.
van Breugel et al. (1986) and Giovannini et al. (1988) have reported observations
of this source with the VLA A-array. These subarcsec-resolution images show a
compact core and a radio jet towards the south-west with a prominent knot in the jet 
separated from the core by $\sim$3 arcsec (3.7 kpc). With our relatively coarse 
resolution of a few arcsec the core flux density is contaminated by emission from the jet. 
From the high-resolution observations the core flux density is 
10.4 mJy at 4873 MHz and 8.2 mJy at 4860 MHz (van Breugel et al. 1986; Giovannini et al.
1988), although the flux density including the knot is $\sim$16.5 mJy. 

Although our observations are of different resolutions, we have attempted to examine
evidence of variability by making images with an angular resolution of 5.4 arcsec along 
the jet axis, which
is similar to the resolution of the FIRST image, so that contamination by emission from 
the jet is similar at the different frequencies.  A comparison of the flux densities in 2005
suggests that there was a strong outburst between about 1990 and 2005 which is apparent in the core
flux densities at 4860 and also possibly at $\sim$1300 MHz (Fig. 7). It is relevant 
to note that in the DDRG
J1453+3308 Konar et al. (2006) reported evidence of significant variability of the
core flux density at cm wavelengths.  Other examples of cores which have been
monitored and show evidence of variability are 3C338 and B2 1144+352 (Giovannini et al. 1998, 1999;
Schoenmakers et al. 1999). It is interesting to note that both these objects are possible examples of
episodic jet activity (cf. Burns, Schwendeman \& White 1983; Schoenmakers et al. 1999).
It is important to determine from more extensive monitoring whether strong core variability may 
be a common characteristic of sources with renewed jet or nuclear activity even if the cores are 
relatively weak. However, it is also relevant to note that strong core variability and superluminal
motion may be seen in sources inclined at small angles to the line of sight. Such sources are also
expected to have dominant cores due to Doppler boosting of the core flux density. Amongst the sources
discussed here, the arcsec scale core in B2 1144+352 contributes almost 80 per cent of the total 
observed flux density at 1400 MHz, while for the other three objects it contributes $\lapp$5 per 
cent, similar to other galaxies of comparable luminosity (cf. Saikia \& Kulkarni 1994; 
Ishwara-Chandra \& Saikia 1999). This is consistent with the detection of superluminal motion
in B2 1144+352 (Giovannini et al. 1999), while 3C338 exhibits two reasonably symmetric parsec-scale
jets and evidence of subluminal motion (Giovannini et al. 1998). The cores of 4C29.30 and J1453+3308
are weak and have not been studied with mas resolution.  

\begin{table}
\begin{center}
\caption{The radio flux densities of 4C\,29.30.}
\begin{tabular}{r c c c c}
\hline
\multicolumn{1}{c}{Freq.} &\multicolumn{1}{c}{Flux density} &\multicolumn{1}{c}{Err}         &  Ref. &Note \\
\multicolumn{1}{c}{(MHz)} &\multicolumn{1}{c}{(mJy)}        &\multicolumn{1}{c}{(mJy)}       &       &      \\
     (1)                  &        (2)                      &          (3)                   & (4)   & (5)  \\
\hline
     74   &  5393        & 755  &    1 &   \\
     151  &  2753        & 147  &    2 &    \\
     178  &  2220        & 445  &    3 &    \\
     240  &  3541        & 496  &    8 & (a)  \\
     325  &  2170        & 217  &    4 & (a)   \\
     333  &  2647        & 370  &    8 & (a)  \\
     408  &  1763        & 88   &    5 &    \\
     605  &  1465        & 103  &    8 & (a)  \\
     1287 &   727        & 51   &    8 & (a)  \\
     1400 &   738        & 30   &    6 &    \\
     1400 &   731        & 37   &    7 & (a)   \\
     1400 &   756        & 23   &    8 & (a)   \\
     2700 &   430        & 10   &    9 &    \\
     4850 &   266        & 40   &   10 &    \\
     4850 &   239        & 24   &   11 &   \\
     4850 &   269        & 35   &   12 &   \\
     4850 &   271        & 14   &    8 & (a)  \\
\hline
\end{tabular}
\end{center}

References along with the names of some of the well-known surveys.
(1) VLSS: VLA Low-frequency Sky Survey;
(2) 7C: Riley et al. 1999; 
(3) 4C: Pilkington \& Scott 1965;
(4) WENSS: Rengelink et al. 1997; 
(5) B2: Colla et al. 1970;
(6) White \& Becker 1992; 
(7) NVSS: Condon et al. 1998; 
(8) this paper; 
(9) Bridle et al. 1977;
(10) Becker, White \& Edwards 1991; 
(11) Langston et al. 1990;
(12) Gregory \& Condon 1991. \\
Notes: (a) The diffuse extended emission  
is visible in these images (Figs. 1, 2 and 3), although the
entire flux density is possibly seen only in the VLA D-array image at 
1400 MHz (Fig. 1) and the Effelsberg image at 4850 MHz (Fig. 2).  
\end{table}


\begin{figure}
\vbox{
   \psfig{file=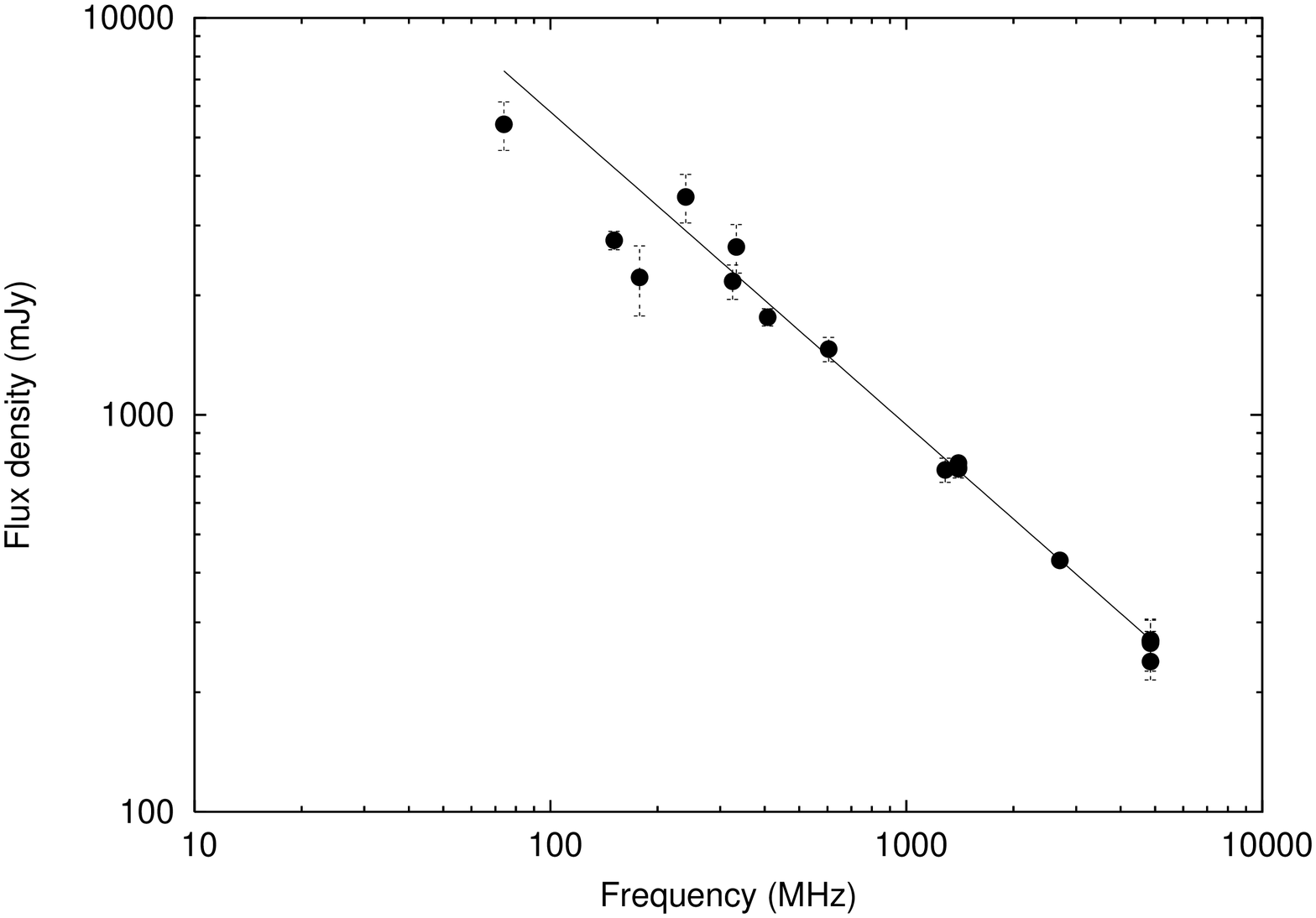,width=3.3in,angle=-90}
   \psfig{file=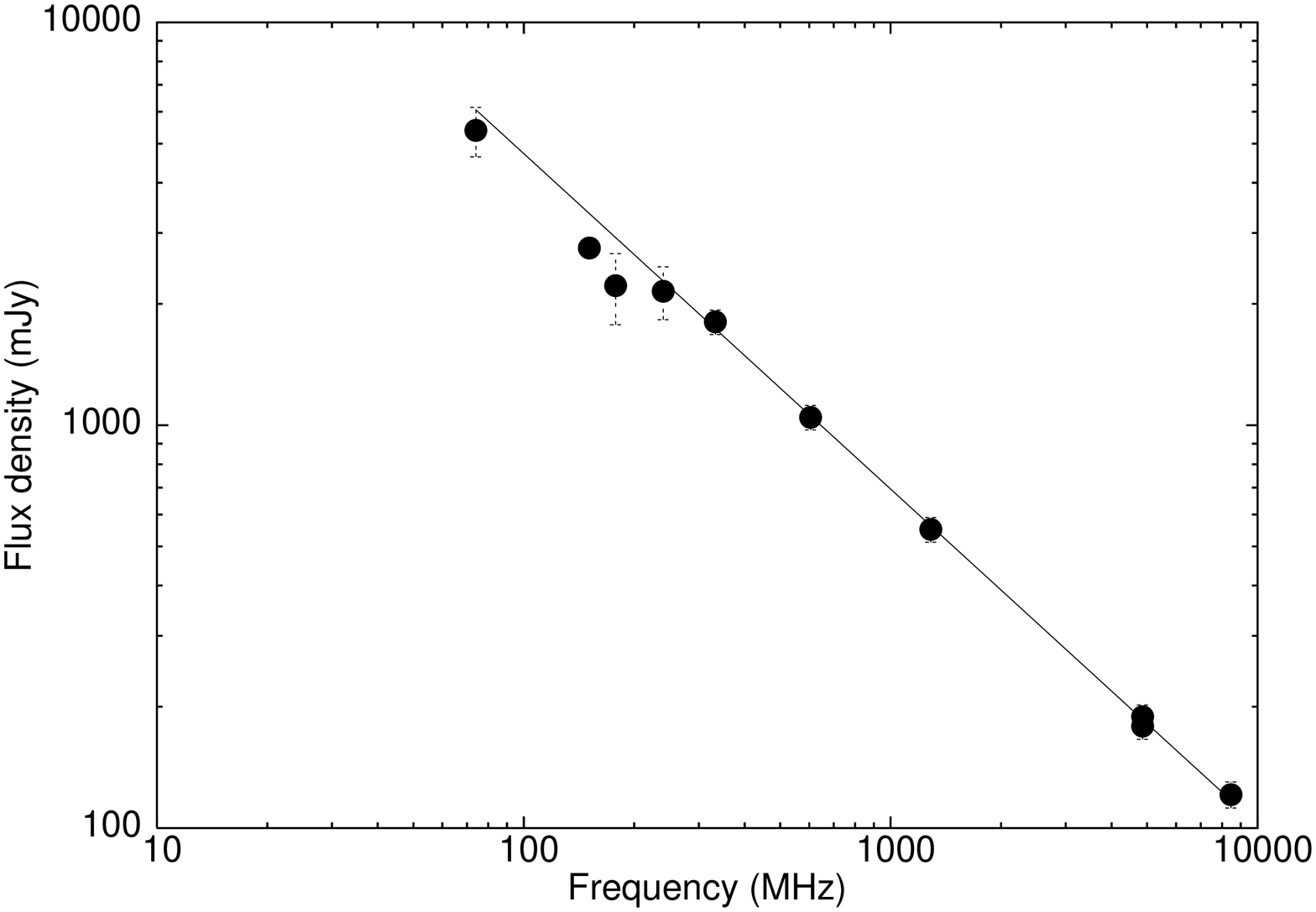,width=3.3in,angle=-90}
}
\caption[]{Upper panel: Spectrum of the 4C29.30 using the flux densities listed in Table 4. 
The linear least-squares fit has been made using measurements only between 240 and 4850 MHz.
Lower panel: Spectrum of the inner double using our GMRT and VLA images between 240 and 8460
MHz. The contributions of the core component within a region of $\sim$4 arcsec (4.9 kpc) have
been subtracted at frequencies $\gapp$1300 MHz. The spectrum shows the low-frequency measurements 
at 74, 151 and 178 MHz listed in Table 4, although the linear least squares fit has been made 
using measurements only between 240 and 8460 MHz. This fit has been extrapolated to lower frequencies.}
\end{figure}

\subsection{Spectra}
The flux densities of the source, 4C29.30, from the literature as well as from our
images which show the extended structure (Figs. 1, 2 and 3) are summarised in Table 4 which
is self explanatory. All the flux densities are consistent with the scale of Baars et al.
(1977) and where necessary have been converted to this scale using the conversion
factors listed by K\"{u}hr et al. (1981). The spectrum obtained from these measurements 
is presented in Fig. 8 (upper panel). There are some caveats which need to be borne 
in mind while examining this spectrum. For the low-frequency measurements where the
diffuse extended emission is more likely to make a significant contribution the noise
levels are often very high and the diffuse emission may not be `visible' in these 
observations. This can be seen clearly in the VLSS image ({\tt http://lwa.nrl.navy.mil/VLSS})
 where the rms noise is approximately 80 mJy beam$^{-1}$ (Fig. 9) and there is no evidence of the 
diffuse extended emission. The deconvolved size of this component is 
64$\times$18 arcsec$^2$ along a PA of 51$^\circ$.  The Cambridge 7C image of 4C29.30 at 151 
MHz is towards the edge of the field and may also represent only the flux density of the inner double. 
The spectral index $\alpha$ (S$\propto\nu^{-\alpha}$), determined from the total flux density 
of the source between 240 and 4850 MHz is 0.79$\pm$0.02.

In the lower panel of Fig. 8 we present the spectrum of the inner-double lobed source 
obtained from our GMRT and VLA images after subtracting the contribution of the core
component at frequencies of $\sim$1300 MHz and above. 
The total flux densities have been estimated from images which have been tapered and
weighted to have the same resolution as that of the 240$-$MHz image. The core contribution
is particularly important at 8460 MHz although it does appear to make a significant
contribution at 4860 MHz. Its effect at 1400 MHz is small, its contribution being within
the errors in the flux density.  
Our observations between 240 and 8460 MHz are consistent with a straight spectrum yielding 
a spectral index of 0.83$\pm$0.01 for the inner double. The extrapolation of the spectrum 
to lower frequencies shows that the 74$-$MHz flux density from the VLSS 
is consistent with this spectral index while the Cambridge
measurements at 151 and 178 MHz are marginally below the expected values. We have also
attempted to fit the spectrum of the inner double including the south-western blob of 
emission. It yields similar results with a spectral index of 0.82$\pm$0.02. 

\begin{figure}
\hbox{
    \psfig{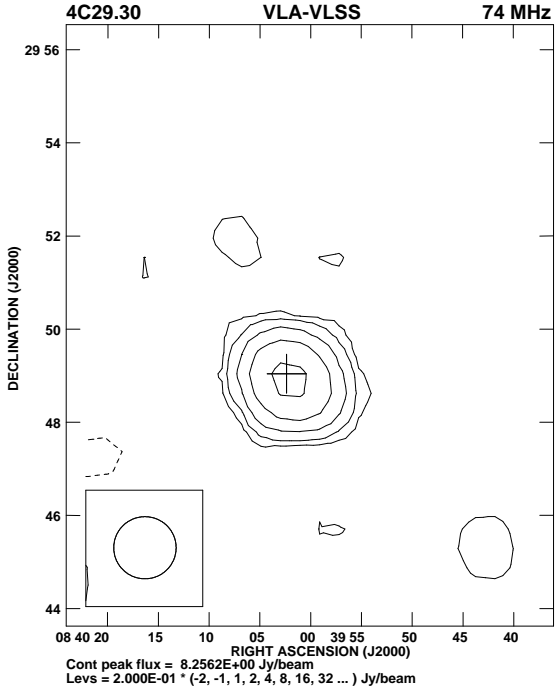}
     }
\caption[]{VLSS image of 4C29.30 with an angular resolution of 80 arcsec showing only
the inner double as a single source. The 
resolution is shown as an ellipse in the bottom left-hand corner while the 
$+$ sign marks the position of the optical galaxy.} 
\end{figure}

\subsection{Polarization properties}
The distributions of the 1400 and 4850~MHz linear polarization of
4C29.30 are shown in Fig. 10. Panels a) and c) show  contours of total intensity, with E-field
polarization vectors superimposed. The lengths of the vectors are proportional to the
polarized intensity. Panels b) and d) show contours of polarized intensity, with
E-field polarization vectors superimposed. Here, the lengths of the vectors are proportional
to the fractional polarization. The integrated polarized flux densities of the
central component  are $41.7\pm2.2$ mJy and $7.9\pm0.5$ mJy at 1400 and 4850~MHz,
respectively. The corresponding degrees of linear polarization are
$6.7\pm0.4$ and $3.3\pm0.3$ per cent. We find different orientations of the electric
E-vectors in different regions of the lobes, as well as differences in the levels of
polarization, suggesting Faraday rotation and depolarization effects. In order to
determine the distribution of the magnetic B-field directions, it is necessary to
correct the radio data for the Faraday rotation measure, and for that purpose
subsequent radio polarisation measurements at other wavelengths with high angular
resolution are needed. At 4850~MHz, the orientation of the E-field in the central
parts of the source agrees with that reported by van Breugel et al. (1986) for
the inner radio lobes (the southern one in particular). Also the large Faraday
rotation measure between 5000 and 1400~MHz observed by van Breugel et al. are
consistent with those indicated by our analysis. This suggests that the polarized
flux is largely from the inner radio structure. However, weak linear polarization is 
also detected from the outer regions far beyond the inner jets and their lobes.

\begin{figure*}
  \psfig{file=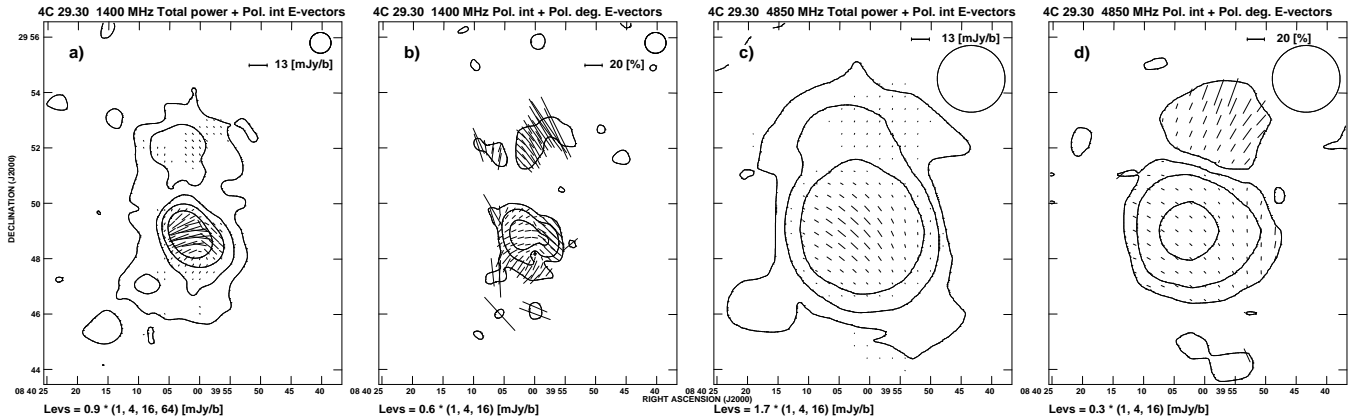,width=7.1in,angle=0}
\caption{Polarization maps  of 4C\,29.30. {\bf a)} The polarization E-vectors at
1400~MHz superimposed on the total-intensity contours and {\bf b)} the
corresponding fractional polarization vectors superimposed on the polarized
intensity contours. {\bf c)} The polarization E-vectors at 4850~MHz superimposed
on the total-intensity contours and {\bf d)} the corresponding fractional
polarization vectors superimposed on the polarized intensity contours.
The size of the beam is indicated by the circles in the right corners of the
images and bars indicate the scale  beams of the polarization vectors.}
\end{figure*}

\section{Discussion}

\subsection{Spectral ageing analysis} 
The spectra using the total flux densities between 240 and 4860 MHz and that of
the inner double using our measurements between 240 and 8460 MHz are consistent
with a single power-law.  We have fitted the spectrum of the inner double after
subtracting the core flux density for the
Jaffe \& Perola (1973; JP), Kardashev-Pacholczyk (KP, Kardashev 1962; Pacholczyk 1970) and
the continuous injection (CI, Pacholczyk 1970) models using the {\tt SYNAGE} package (Murgia 1996).  
The break frequency obtained from these fits are rather large ($\gapp$7$\times$10$^5$ GHz)
and have huge uncertainties because the spectrum is practically straight.  These
conclusions remain unaffected even if the flux density at 74 MHz is included in the analysis.

In order to calculate the global mean magnetic field strength and estimate the
spectral age we have adopted the formula given by Miley (1980), a spheroidal and cylindrical
geometry for the diffuse extended emission and the inner double respectively,
a filling factor of $\sim$1, a proton to electron energy ratio of unity
and lower- and higher-frequency cutoffs of 10 MHz and 100 GHz respectively.
The equipartition magnetic field estimate for the inner double
is 0.43$\pm$0.06 nT, indicating that for the above-mentioned break frequency
the inferred age is $\lapp$0.12 Myr. (For conversion
of the magnetic field to $\mu$G which is also commonly used, 1nT = 10 $\mu$G).   
However, adopting a more conservative break frequency of $\gapp$10 GHz from the
observed spectrum, the spectral age is $\lapp$33 Myr. 

We have attempted to determine the spectrum of
the diffuse extended emission by subtracting the flux density of the inner
double from that of the total emission in the images where this is reliably
seen. For this purpose we have used 
our low-resolution GMRT images at 240, 333 and 605 MHz, our
VLA D-array and NVSS images at 1400 MHz and our Effelsberg measurements at 4850 MHz to
represent the total flux density of the source. Our GMRT image at 333 MHz has a better
rms noise than the WENSS image and gives a larger value of the total flux density. 
After subtracting the 
contribution of the inner double-lobed source, estimated from our GMRT and VLA images
of similar angular resolution of $\sim$12 arcsec, we plot the spectrum of the
diffuse emission in Fig. 11. Although there are significant uncertainties,
a least-squares fit yields a spectral index of 1.26$\pm$0.07. However, given the possibility 
that we are still missing significant amounts of flux density at the lower frequencies, as
discussed earlier, the spectral index could be steeper. Making a very
rough estimate of the possible increase in the flux density by assuming that the diffuse emission 
not detected at the low frequencies compared with the VLA D-array image (Fig. 1) is just below the 
first contour level yields a spectral index of 1.34$\pm$0.05 for the diffuse emission. These points
are indicated by open circles in Fig. 11. The above two values are
consistent within the errors. These estimates suggest that the halo is likely to be much older
than the inner double. 

We have also estimated the spectral index from 240 to 1400 MHz in the vicinity of 
the peak in the northern diffuse emission by considering a box with a size of $\sim$150 arcsec
centred on the peak in the 45-arcsec resolution images. The flux densities at  
240, 333, 605 and 1400 MHz are 188$\pm$26, 139$\pm$19, 87$\pm$6 and 44$\pm$3 mJy respectively.  
A least-squares fit yields a spectral index of 0.81$\pm$0.01 for this region, which is 
similar to the spectral index of the inner double, suggesting that this 
value might be close to the injection spectrum, which is rather steep.
The magnetic field strength in the vicinity of the peak in the northern
diffuse emission is 0.12$\pm$0.02 nT. The straight spectrum of this
region till $\sim$1400 MHz yields a spectral age of $\lapp$100 Myr, 
consistent with an interruption of activity for $\lapp$100 Myr.
For the diffuse extended emission, the spectral break 
is likely to be lower than $\sim$240 MHz, which yields a spectral age 
$\gapp$200 Myr. To get a reliable value of the spectral break and 
estimate the age using {\tt SYNAGE} we need more sensitive images
at low frequencies, especially below $\sim$240 MHz.  

\begin{figure}
\vbox{
    \psfig{file=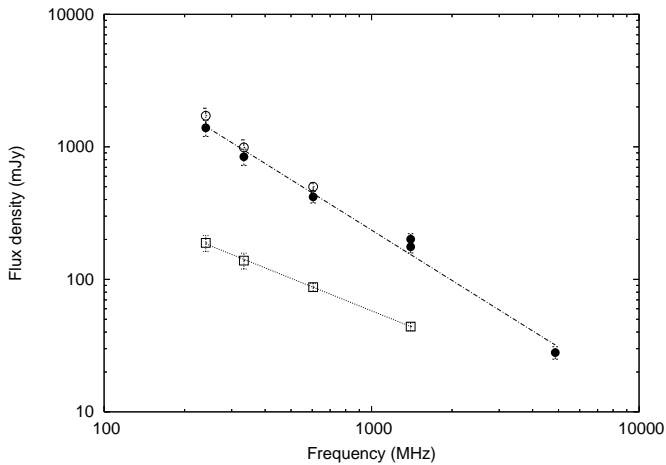,width=3.5in,angle=-90}
     }
\caption[]{The radio spectrum of the diffuse emission. Filled circles denote the
flux density of the diffuse extended emission after subtracting the
contribution of the inner double-lobed source, the latter being estimated from our 
GMRT and VLA images of similar angular resolution of $\sim$12 arcsec. The open circles denote the flux
density after incorporating a rough estimate of the flux density not seen in 
the GMRT low-frequency images compared with the VLA D-array image. The open squares
denote the flux densities in a box with a size of $\sim$150 arcsec centred in the
peak of the northern diffuse emission. The dashed lines represent the least-square fits
to the points. For the diffuse emission this has been done considering only the
points represented by the filled circles.}
\end{figure}
  
Nevertheless, while interpreting these numbers caveats related to the evolution of
the local magnetic field in the lobes need to be borne in mind (e.g. Rudnick, Katz-Stone
\& Anderson 1994; Jones, Ryu \& Engel 1999; Blundell \& Rawlings 2000).
While Kaiser, Schoenmakers \& R\"{o}ttgering (2000) have suggested that spectral 
and dynamical ages are comparable if
bulk backflow and both radiative and adiabatic losses are taken into account in a
self-consistent manner, Blundell \& Rawlings (2000) suggest that this may be so only
in the young sources with ages much less than 10 Myr. More recently Machalski et al.
(2006) have examined the dynamical ages of FRII radio sources and find that while 
these agree with the spectral ages for objects less than 10 Myr, their method may be
applicable for older sources as well. For all objects they find realistic jet advance
velocities.  

\subsection{X-ray observations of the inner double}
The existing Chandra data for 4C29.30 obtained on 2001 April 8 (OBSID=2135,
previously analyzed by Gambill et al. 2003 and Sambruna et al. 2004) reveal x-ray
emission associated with the inner radio structure. Although
the photon statistics from the archival 7.5 ksec exposure is very low, several
x-ray components are visible, namely a highly obscured nucleus, a diffuse
galactic-scale halo, hotspot (`HS'), counter-hotpsot (`C-HS'), and possibly
both jets connecting hotspot and counter-hotspot with the nucleus. There
is also a possible x-ray excess to the south-west from the nucleus, co-spatial
with a region of enhanced optical-line emission.

We reprocessed the archival Chandra data using CALDBv.3. The 7.7-ksec ACIS-S observation
detected a total of 123$\pm$11 counts from the core (a circle of 1.5 arcsec radius). Most
of these counts (105) exceed 2-keV energies, which indicate a very hard spectrum. The
absorbed power-law model fit shows strong evidence for large intrinsic absorption of
$N_{\rm H}\sim(3.5\pm0.9)\times10^{23}$ cm$^{-2}$ for the fixed photon index
of $\Gamma=1.7$, implying the intrinsic unabsorbed luminosity of $L_{\rm 2-10 \,keV}
\approx 3\times10^{43}$ erg s$^{-1}$. This is in agreement with the results by Gambill et al.
(2003), although our number of total detected counts is larger because we used the 
updated Chandra calibration to process the data. Such high values of column density
have been reported for some compact radio sources (e.g. Guainazzi et al. 2006;
Vink et al. 2006). We note that the 
nucleus of 4C29.30 reveals optical features typical of Seyfert 2 and
narrow-line radio galaxies (van Breugel et al. 1986).

Both hotspot and the counter-hotspot regions are pronounced in x-rays
(Sambruna et al. 2004).  Note that
low-power radio galaxies like 4C~29.30 usually do not exhibit strong
shocks at the terminal regions of the jets.  The Chandra data
indicate significant differences in the spectra of both features.  In
particular, in the case of the hotspot the number of soft ($0.5-2$
keV) and hard ($2-10$ keV) total counts is $S=16$(B=1), where B 
indicates background counts, and $H=6$(B=1), respectively, giving the
hardness ratio $S/H = 2.9^{+3.9}_{-1.3}$; in the case of the
counter-hotspots the analogous numbers are $S=9$(B=1), $H=1$(B=1), and
$S/H = 10.5^{+11.3}_{-6.5}$ (see Park et al. 2006 for the description
of the Bayesian methods we used to calculate the hardness ratio and errors).  
This suggests that the observed hotspot-related x-ray emission, especially
in the counter hotspot, may be a mixture of the thermal and non-thermal radiation.    
The thermal scenario may be particularly
relevant for the counter-hotspot region, which is in fact co-spatial with a zone 
of strong optical
lines emission. It is interesting to note that in the radio galaxy Cen A, sharing
many morphological similarities with the object discussed here,
Kraft et al. (2003) reported detection of thermal x-ray emission at the counter-jet
termination region due to galactic gas compressed by the expanding radio outflow.
We suggest that the Chandra
data may indicate similar jet/ambient medium interactions in the counter hotspot of
4C29.30, although deeper observations are needed to confirm such a scenario.
In this context it is relevant to note that the line-emitting gaseous filaments
extending along the inner radio structure seem to be heated and
accelerated by the expanding jet rather than photoionized by the ambient
medium (van Breugel et al. 1986). Evidence of similar jet-cloud interactions
have been reported for many radio sources such as the compact steep-spectrum sources 
3C48 (Gupta, Srianand \& Saikia 2005) and 3C67, 277.1, and 303.1 (O'Dea et al. 2002).

Another interesting constraint is offered by the polarization studies. As
discussed previously, the polarized flux-density  from the central parts
of 4C\,29.30 is most likely dominated by the inner lobes. As argued by van
Breugel et al. (1986), large differences between the orientations of the
E-vectors in this region when observed at 1400~MHz and 4850~MHz suggest
Faraday rotation effects, most likely not due to the observed optical
emission-line gaseous clouds, but rather due to the `intercloud medium'.
However, similar rotation of the electric field vector is observed by us
also in the outer regions, far ($> 100$\,kpc) beyond the inner radio
structure studied by van Breugel et al. This may suggest that the Faraday
screen should not be identified with the gaseous medium co-spatial with
the inner ($\sim$ few kpc-scale) optical line-emitting clouds, but rather
with the ambient thermal gas on larger scales, pushed-out and compressed by the
expanding fossil halo, eventually partly mixed with the plasma of the lobes.

\section{Concluding remarks}
We present the results of multifrequency radio observations of the
radio galaxy 4C29.30 using the GMRT, the VLA and the Effelsberg telescope. 

\begin{enumerate}
\item The low-resolution images with the VLA, GMRT and the Effelsberg telescope show 
evidence of a large scale diffuse emission with an angular scale of $\sim$520 arcsec
(639 kpc) in addition to the small-scale inner double which has an angular size
of $\sim$29 arcsec (36 kpc). The structure of the inner double is
similar to that of FRII radio sources although its radio luminosity is in the FRI
category. This is similar to some of the inner doubles in the DDRGs.

\item The GMRT and VLA observations of the inner double show that 
it has a spectral index of $\sim$0.8 with no evidence of curvature in 
its spectrum from 240 to 8460 MHz, suggesting that the inner double which has an 
edge-brightened structure is young. 
We have fitted the spectrum using {\tt SYNAGE} for the JP, KP and CI models.
The break frequency is large ($\gapp$10$^6$ GHz) but this value has huge uncertainties
because the spectrum is practically straight. 
Its equipartition magnetic field is 0.43$\pm$0.06 nT using the classical formula.
The inferred spectral age for the inner double for a conservative lower limit of
$\sim$10 GHz for the break frequency is $\lapp$33 Myr.
 
\item The diffuse outer emission represents emission from an earlier cycle of activity. 
Its total spectrum is steep with a spectral index of $\sim$1.3, with the break
frequency being $\lapp$240 MHz. 
Using the classical formula the equipartition magnetic field is
0.07$\pm$0.01 nT which yields a radiative age of $\gapp$200 Myr. 
  
\item The spectral index for the inner double as well as in the vicinity of the northern
peak of the diffuse extended emission is $\sim$0.8, suggesting that the injection
spectral index is close to this value. Although the injection spectrum is somewhat steep, 
the similarity for the inner double and the outer diffuse emission is reminescent
of J1453+3308 where the outer and inner doubles have similar injection 
spectral indices (Konar et al. 2006). 

\item The magnetic field in the vicinity of the northern peak of the diffuse extended 
emission is 0.12$\pm$0.02 nT. Given its straight spectrum between 240
and 1400 MHz, this yields a spectral age of $\lapp$100 Myr. Although higher frequency observations
are required to extend the spectrum for both the inner double and the diffuse extended emission, 
the present data suggests a
time scale of interruption of activity of $\lapp$100 Myr. However, the known caveats regarding
estimation of spectral ages should be borne in mind.
 
\item The radio core exhibits evidence of variability. This is
similar to that of the DDRG J1453+3308, suggesting that significant core variability
may often occur in galaxies with evidence of recurrent activity.

\item The hotspots of the inner double have been detected at x-ray and optical wavelengths
using the Chandra x-ray Observatory and the Hubble Space Telescope. Our reanalysis of the
x-ray data suggests that the counter hotspot consists of both thermal and nonthermal material. 
The thermal x-ray emission may be due to hot gas compressed by the expanding radio-emitting outflow
on kpc scales.
\end{enumerate}

\section*{Acknowledgments} 
We thank the anonymous referee for his valuable comments which made us re-look at our data,
and present a more satisfactory interpretation of the results. We also thank Bill Cotton 
and Ger de Bruyn for clarifications on the VLSS and WENSS flux density 
scales, and Julia  Riley and Dave Green for information on the 7C images and flux density scale.
We thank the staffs of GMRT,  VLA and the Effelsberg telescope for their help with the
observations. The GMRT is a national facility operated by the
National Centre for Radio Astrophysics of the Tata Institute of Fundamental Research. 
The National Radio Astronomy Observatory  is a facility of the National Science Foundation
operated under co-operative agreement by Associated Universities Inc. The Effelsberg 100-m
telescope is operated by the Max-Planck-Institut f\"ur Radioastronomie (MPIfR).  
This research has made use of the NASA/IPAC extragalactic database (NED) which is operated
by the Jet Propulsion Laboratory, Caltech, under contract with the National Aeronautics
and Space Administration. We acknowledge use of the  Digitized Sky Surveys which were produced 
at the Space Telescope Science Institute under U.S. Government grant NAG W-2166. The images of 
these surveys are based on photographic data obtained using the Oschin Schmidt Telescope on Palomar 
Mountain and the UK Schmidt Telescope.  
We thank Matteo Murgia for access to the {\tt SYNAGE} software and useful discussions.
MJ was partly supported by the Polish State funds for scientific research in years
2005--2007 under contract No. 0425/PO3/2005/29.
{\L}S was supported by  MEiN through the research project
1-P03D-003-29 during the years 2005-2008.  This research was funded in part by NASA
contract NAS8-39073: partial support for this work (AS and {\L}S) was
provided by the  National Aeronautics and Space Administration through
Chandra Award Number GO5-01164X and issued by the Chandra X-Ray Observatory
Center, which is operated by the  Smithsonian Astrophysical Observatory
for and on behalf of  NASA under contract NAS8-39073.

{}

\end{document}